# Model for processive movement of dynein

**Ping Xie, Shuo-Xing Dou, and Peng-Ye Wang**

*Laboratory of Soft Matter Physics, Beijing National Laboratory for Condensed Matter Physics, Institute of Physics, Chinese Academy of Sciences, Beijing 100080, China*

A model for the processive movement of dynein is presented based on experimental observations available. In the model, the change from strong microtubule-binding to weak binding of dynein is determined naturally by the variation of the relative orientation between the two interacting surfaces of the stalk tip and the microtubule as the stalk rotates from the ADP.Vi-state orientation to the apo-state orientation. This means that the puzzling communication from the ATP binding site in the globular head to the MT-binding site in the tip of the stalk, which is prerequisite in the conventional model, is not required. Using the present model, the previous experimental results, such as (i) the step size of a dynein being an integer times of the period of the MT lattice, (ii) the dependence of the step size on load, *i.e.*, the step size decreasing with the increase of load, and (iii) the stall force being proportional to [ATP] at low [ATP] and becoming saturated at high [ATP], are well explained.

## 1. INTRODUCTION

Dyneins are microtubule-based motor proteins, which fall broadly into two principle classes: axonemal dynein and cytoplasmic dynein (Gibbons 1998; Lodish *et al.* 2000). Axonemal dynein was the first to be discovered and functions as a molecular engine for ciliary and flagellar movement (Gibbons 1963). More than 20 years after the first discovery of the axonemal dynein, cytoplasmic dynein was identified (Paschal *et al.* 1987) and found to be involved in transport of organelles and vesicles as well as in spindle assembly and chromosome segregation (Hirokawa 1998; Wittmann *et al.* 2001). Each dynein is a complex of between 1 and 3 heavy chains, each with a relative molecular mass greater than 500,000, together with a number of intermediate and light chains. Each heavy chain constitutes the fundamental motor unit. Electron microscopy has established that the heavy chain folds to form a globular head with two elongated structures, the stalk and stem, emerging from it. The stalk and the stem bind the microtubule track and cargo, respectively. The stalk is most probably an anti-parallel coiled-coil, with a small microtubule-binding domain at its tip.

With extensive investigations using different experimental methods, such as biochemical, biophysical, and single-molecular approaches, many dynamical behaviors of dyneins *in vitro* have been gradually elucidated. Important mechanical properties such as stall force, step size and velocity have been determined systematically (Shingyoji *et al.* 1998; Sakakibara *et al.* 1999; King & Schroer 2000; Hirakawa *et al.* 2000; Mallik *et al.* 2004). However, the microscopic mechanism of its processive movement is still not very clear. Generally, two types of models have been proposed. One type are the thermal ratchet models in which the motor is viewed as a Brownian particle moving in two (or more) periodic but spatially asymmetric stochastically switched potentials (Astumian 1997 ; Jülicher et al. 1997). Another type are the power-stroke models that are presented based on structural observations (Burgess *et al.* 2003; Mallik *et al.* 2004). In these models, the power stroke is generated by the relative rotation between the stalk and the stem when dynein makes a transition from ADP.Vi-state to apo-state. In order to have a unidirectional movement, it has to be assumed that ATP binding to the globular head reduces the MT-binding affinity. This means that a communication should exist between the ATP-binding site in the globular head and the MT-binding site in the tip of the stalk, which, however, is difficult to imagine in view of the dynein structure.

In this paper we present another power-stroke model for the unidirectional movement of dynein based on the available experimental observations. Our model is different from the conventional power-stroke models in that, in our model, the change between strong and weak MT-bindings of the tip during the nucleotide-state transitions does not need to have the communication between the two sites, but is resulted naturally from the varying orientation of the stalk with respect to MT during the stalk rotation. Using our model, the step size of a cytoplasmic dynein being an integer times of the period $d = 8$ nm of the MT lattice and the step size decreasing with the increase of load can be well explained. In particular, the theoretical results that the stall force is proportional to [ATP] at low [ATP] and becomes saturation at high [ATP] are in good agreement with previous experimental results.

## 2. MODEL

We make the following three assumptions:

1. *Strong MT-binding of the stalk tip promotes a conformational change in the active site, thus activating release of the ATP hydrolysis products: ADP and phosphate.*

This assumption is consistent with experimental results on activation of the ATPase by MT (Omoto & Johnson 1986), where it was shown that MT has negligible effect on ATP-binding rate whereas it can enhance greatly the product-release rate, *e.g.*, by about eleven times for *Tetrahymena* 22S dynein. This can be understood as follows. When the stalk tip is in the strong MT-binding state, the tip remains in a fixed orientation relative to MT, similar to the cases of kinesin head binding strongly to MT (Mandelkow & Hoenger 1999, Xie *et al.* 2004) and myosin head binding strongly to actin filament (Volkman & Hanein 2000, Xie *et al.* 2004). This fixed orientation of the stalk tip can result in an internal force (or torque) in the stalk that is connected with the globular head. This force (or torque) may thus lead to the conformational change in the active site located within the globular head.

2. *There are two nucleotide-state-dependent conformations of dynein as shown in figures 1a and 1b.*

In another word, the release of ADP and $P_i$ leads to the rotation of the stalk from the position as shown in figure 1*a* to that as shown in figure 1*b*, and ATP binding has the opposite effect: leading to the rotation of the stalk from that as shown in figure 1*b* to that as shown in figure 1*a*. This assumption is supported by structural study of the inner-arm dynein c of *Chlamydomonas* flagella by using electron microscopy (Burgess *et al.* 2003).

3. *The stalk can be considered as rigid* (Gee *et al.* 1997).

### A. Axonemal Dyneins

Based on the above assumptions we propose a model for the unidirectional and processive movement of MT by an axonemal dynein motor, as observed in the *in vitro* MT sliding studies by Sakakibara *et al.* (1999) where the dynein is fixed and MT can be moved. As dynein may bind to MT in two different nucleotide states or conformational states, *i.e.*, ADP.Vi-state and apo-state, at the onset, we consider the two cases separately.

**(i)** Dynein binds strongly to MT in ADP.Vi state (figure 2*a*). In this case, dynein releases ADP and $P_i$ with a high rate. This leads to the rotation of the stalk, thus driving the MT to the positive end by a distance $L_S$ (figure 2*b*). Since the stalk can be considered as rigid (Gee *et al.* 1997), it is clear that, as the stalk rotates, the gap between the tip and MT is enlarged and thus the MT-binding strength of the tip decreases. In another word, the tip changes from strong MT-binding at the beginning of the stalk rotation to weak MT-binding at the end of the rotation. Then after ATP binding to dynein, the stalk rotates back and the dynein returns to its original conformational state as shown in figure 2*a*. In this process the MT may also moves back due to the weak MT-binding of the tip and/or existence of load, but it does not return to the original position as shown in figure 2*a* (see discussion below). Now the tip binds strongly again to MT at a new binding site. After ATP hydrolysis, a chemical cycle is finished and a mechanical stepping of MT is completed (figure 2*c*). Note that the step size will be *n* (*n* is an integer) times of the period *d* = 8 nm of the MT lattice provided that $(n-1/2)d < L_S - L_M < (n+1/2)d$, where $L_S$ is the movement distance of the tip before slipping of the tip relative to MT occurs during the rotation of the stalk as dynein changes from ADP.Vi state to apo state, and $L_M$ is the relaxation distance of MT (not shown in figure 2*c*) when the stalk returns to its original orientation.

Here we give an example to illustrate the above description in detail. For simplicity, it is assumed that the temporal evolution of the rotation velocity, *V*, of the stalk has a form as shown by the blue line in figure 3. By integration we can obtain the temporal evolution of the rotation angle. Assuming that the binding force of the tip to MT is inversely proportional to the rotation angle, we thus have the temporal evolution of the binding force, $F_b$, with a form as shown by red or green lines in figure 3, where the red line represents the rotation from the ADP.Vi-state to apo-state and the green line from the apo-state to ADP.Vi-state. From Stokes law we have the following results: As the stalk rotates, once $F_b$ becomes smaller than $F_s + F_{load}$, the slipping of the tip relative to MT occurs, where $F_s$ is the Stokes force on MT (with a form shown by black line in figure 3 as calculated by using the moving velocity of the stalk tip) and $F_{load}$ is the force exerted on the MT. Therefore, from figure 3 we see that, as the stalk rotates from the ADP.Vi to apo orientations, $F_b$ is always larger than $F_s$ and thus under zero load the slipping does not occur; whereas, as the stalk rotates from the apo to ADP.Vi orientations, $F_b$ becomes smaller than $F_s$ and thus the slipping begins to occur near the beginning of the rotation (at the point marked by a arrow). When slipping occurs the motion of MT can be considered as determined solely by the Langevin noise and load.

**(ii)** Dynein binds strongly to MT in apo-state (figure 2*a'*). Then after ATP binding, the stalk rotates from the apo-state orientation to the ADP.Vi-state orientation (figure 2*b'*). This drives the MT moving a distance $L_S$ to the minus end and, at the same time, the tip changes from strong MT-binding to weak MT-binding states. At present, due to the weak MT-binding dynein releases ADP and $P_i$ with a very low rate and, within the long-time period of release, the binding of the tip to the binding site (I) can change from the weak binding to the strong binding, as shown in Fig. 2(c'). This becomes the same as that shown in Fig. 2(a). Thus, except for the first step (from figures 2*a'* to 2*c'*), MT will then move to the positive end processively with the step size of *nd* just as the same as in Case (i).

## B.  Cytoplasmic Dyneins

For the case that MT is fixed and dyneins can move, as for the cytoplasmic dynein and in the case of the experiments in King & Schroer (2000) and Mallik *et al.* (2004), we describe the processive movement of single dynein molecules along MT as follows.

As discussed before, there exist two conformational states for a dynein binding to MT, *i.e.*, ADP.Vi state and apo state. Thus we consider the following two cases separately.

**(i)** Dynein binds strongly to MT in ADP.Vi state (figure *4a*). In this case, dynein releases ADP and $P_i$ with a high rate. This leads to the rotation of the stalk, thus driving the head to move toward the MT minus end by a distance $L_S$ (figure 4*b*). Here we also refer to figure 3, where, however, $F_s$ is now the Stokes force on the dynein head calculated by using the moving velocity of the head. When there is no load, slipping of the tip relative to MT does not occur as the stalk rotates from figures 4*a* to 4*b* because $F_b$ (red line in figure 3) is always larger than $F_s$. Then after ATP binding to dynein, the stalk rotates and returns to the original state (figure 4*c*). However, slipping occurs near the beginning of the rotation. Therefore, a chemical cycle is completed with a mechanical step of size *nd* if $(n-1/2)d < L_S - L_H < (n+1/2)d$, where *n* is an integer, *d* is the period of the MT lattice, $L_S$ is the movement distance of the head during the rotation of the stalk as dynein changes from ADP.Vi state to apo state and $L_H$ is the relaxation distance of the head due to the weak MT-binding of the tip before the slipping. (Note that, since the size of the head is much larger than that of the tip, as the stalk rotates the movement of the head can be negligible due to the much larger drag force when the dynein is free from external force.) When there is a load, the slipping of the tip relative to MT begins to occur near the end of the dynein conformational change from ADP.Vi state to apo state, and thus as the dynein returns from apo state to ADP.Vi state the slipping occurs at the beginning of stalk rotation. Therefore, a chemical cycle is finished with a step size of *nd*, where *n* is an integer which satisfies $\frac{L_{S1} - L_{H1}}{d} - \frac{1}{2} < n < \frac{L_{S1} - L_{H1}}{d} + \frac{1}{2}$, $L_{S1}$ is the movement distance of the head before slipping occurs during one rotation of the stalk and $L_{H1}$ is the relaxation distance of the head due to the load. It is noted that, as the load is increased, $L_{S1}$ decreases and $L_{H1}$ increases. Thus we have the following results: *The step size decreases with the increase of the load*. This is consistent with the experimental observations (Mallik *et al.* 2004). When the step size decreases to zero the corresponding load is the stall force. Similarly, from our above discussion we can predict that, when a low load is applied to pull the movement of dynein, its step size increases. However, as seen from figure 3, for a large negative load, slipping never occurs during either stalk rotation from the ADP.Vi-state to apo-state orientations or that from apo-state to ADP.Vi-state orientations, thus the dynein cannot move processively.

The above discussion is related to the case of saturating ATP concentration. In the following we will discuss on the behavior of stall force at low ATP concentration.

First, we study the case at extremely low [ATP]. The dynein binds strongly to MT in ADP.Vi state as shown in figure 5a. After release of ADP and $P_i$, the dynein binds weakly to MT in apo-state as shown in figure 5b. Since the ATP concentration is extremely low, the dynein will remain in apo-state for a long time. Within this long-time period, the stalk tip can change from weak MT-binding to strong MT-binding as shown in figure 5c even without external force. Then after ATP binding, the dynein changes to the state as shown in figure 5d. Now dynein binds weakly to MT and thus releases ADP and $P_i$ with a very low rate and, and within the long-time period of product release, the stalk tip can change from weak MT-binding to strong MT-binding. Thus the dynein returns to its original state and position (figure 5a). Therefore, at extremely low ATP concentration the stall force is zero.

With increasing [ATP], ATP-binding time $t_b$ becomes shorter. But from figure 5 it can be seen that a load can accelerate the change of the apo-state dynein from weak MT binding (figure 5b) to strong MT-binding (figure 5c). Thus if the load is large enough so that the time $t_0$ required for the change from figures 5b to 5c satisfies $t_0 = t_b$, dynein goes to the state as shown in figure 5c without ATP binding. Thus dynein still remains stalled as in the case of extremely low [ATP]. The load required to stall the dynein movement along MT corresponds to the stall force $F_{stall}$. From the above analysis it can be seen that $F_{stall}$ should increase with [ATP]. Here we give an approximate description of this relation (See Appendix A for a more correct description). Under an over-damped condition, the rotation velocity, $\omega$, of the dynein head is approximately proportional to the exerted torque which is in turn proportional to the exerted force $F_{stall}$. Thus we have

$$t_0 = \theta_0 / \omega = C_1 / F_{stall}, \tag{1}$$

where $C_1$ is a constant. Since

$$t_b = C_2/[\text{ATP}], \tag{2}$$

where $C_2$ is a constant, from equations (1) and (2) and using $t_0 = t_b$ we obtain

$$F_{stall} = C_3[\text{ATP}], \tag{3}$$

where $C_3 = C_1/C_2$ is a constant. Equation (3) means that, at low [ATP], the stall force is proportional to [ATP].

At very high [ATP], the stall force should correspond to the load by which the head is moved back to the position as shown in figure 5*a* from that as shown in figure 5*b* during the period of the rotation of the stalk from the apo-state to ADP.Vi-state, because ATP binding occurs before the completion of this stalk rotation. Thus the stall force is approximately constant at very high [ATP]. Here we give an estimate of the magnitude for this saturating stall force: Under an over-damped condition, $F_{stall} \approx \frac{\Gamma L_{S1}}{t_0}$. From Stokes formula $\Gamma = 6\pi\eta r_k \approx 4 \times 10^{-10}\,\text{kg s}^{-1}$, where the viscosity $\eta$ of the aqueous medium of a cell around the dynein head is approximately $0.015\,\text{g cm}^{-1}\,\text{s}^{-1}$ (Swaminathan *et al.* 1997) and the dynein head is approximated as a sphere with radius $r_k \approx 15\,\text{nm}$. Since the experimental value of the rotational time of the stalk is unavailable, we take the rotational time the same order as that of the neck of myosin, *i.e.*, 10 μs (Roopnarine *et al.* 1998). Taking $L_{S1} = 24\,\text{nm}$ we have $F_{stall} = 0.96\,\text{pN}$ which is very close to the experimental result of 1.1 pN. The above conclusion that *the stall force is proportional to [ATP] at low [ATP] and becomes saturating at high [ATP]* is in good agreement with the experimental results [figure 1(b) in Mallik *et al.* (2004)]. It is emphasized that this conclusion is different from that derived from the conventional model (Burgess *et al.* 2003), where it is assumed that ATP binding leads to weak MT-binding of the tip and then it can be derived that *the stall force should be independent of [ATP]*.

(ii) At the beginning, dynein binds strongly to MT in apo state. Similar to the case of axonemal dynein, except for the first step, the cytoplasmic dynein will move processively to the minus end of MT, same as that in case (i).

## 3. CONCLUSION

Based on the previous available experimental observations, such as (i) strong binding of MT activating release of ADP and phosphate, (ii) existence of two conformational states, i.e., ADP.Vi-dynein and apo-dynein, and (iii) a rigid stalk, we present a model for the processive movement of dynein. Different from the conventional models, where it is assumed that the ATP binding leads to the weak binding of the tip to MT (*i.e.*, the communication from the ATP binding site in the globular head to the MT-binding site in the tip of the stalk must be required), our model does not need this puzzling communication and the change from the strong MT binding of the tip to the weak binding is determined naturally by the varying relative orientation between the two interacting surfaces as the stalk rotates from the ADP.Vi-state orientation to the apo-state one. Using the model, (i) the step size of a dynein being an integer times of the period of the MT lattice is explained; (ii) the dependence of the step size on load, *i.e.*, the step size decreasing with the increase of load is explained; (iii) the stall force being proportional to [ATP] at low [ATP] and becoming saturated at high [ATP] is explained.

A remarkable prediction of our model is that, for a load pulling the movement of dynein, the step size increases. However, for a large pulling load, the dynein cannot move processively. Thus it is interesting to experimentally verify this prediction in the future.

This work was supported by the National Natural Science Foundation of China.

# APPENDIX A: PRECISE FORMULA FOR THE [ATP] DEPENDENCE OF STALL FORCE AT LOW [ATP]

Consider the over-damped head rotating around the tip of stalk, as shown in figure 5(b). The rotation equation is

$$\gamma \frac{d\theta}{dt} = T_{stall} + \xi(t), \tag{A1}$$

where $\gamma$ is the frictional drag coefficient, $\theta$ the rotational angle, $T_{stall}$ the exerted torque which can be approximately written as $T_{stall} = F_{stall} l$, with $l$ the stalk length, and $\xi(t)$ is the Brownian torque due to thermal fluctuations which satisfies $\langle \xi(t) \rangle = 0$ and $\langle \xi(t)\xi(t') \rangle = 2k_B T \gamma \delta(t-t')$. Solving equation (A1) we obtain the mean first-passage time $t_0$ for rotation of angle $\theta_0$ as follows

$$t_0 = \frac{\gamma}{F_{stall} l} \left\{ \theta_0 + \frac{k_B T}{F_{stall} l} \left[ \exp(-\frac{F_{stall} l}{k_B T} \theta_0) - 1 \right] \right\}. \tag{A2}$$

On the other hand, the ATP binding time can be written as

$$t_b = const./[\text{ATP}]. \tag{A3}$$

Thus from equations (A2) and (A3) and using $t_0 = t_b$, the relation between the stall force $F_{stall}$ and the ATP concentration is obtained as follows

$$\frac{\gamma}{F_{stall} l} \left\{ \theta_0 + \frac{k_B T}{F_{stall} l} \left[ \exp(-\frac{F_{stall} l}{k_B T} \theta_0) - 1 \right] \right\} = \frac{const.}{[\text{ATP}]}. \tag{A4}$$

**FIGURES**

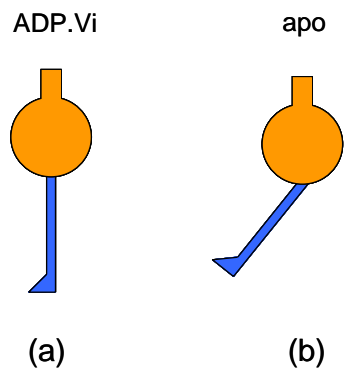

Figure 1. The two different nucleotide-state-dependent conformations of dynein. The globular head is in orange and the stalk in blue. The tip of the stalk is represented by a triangle.

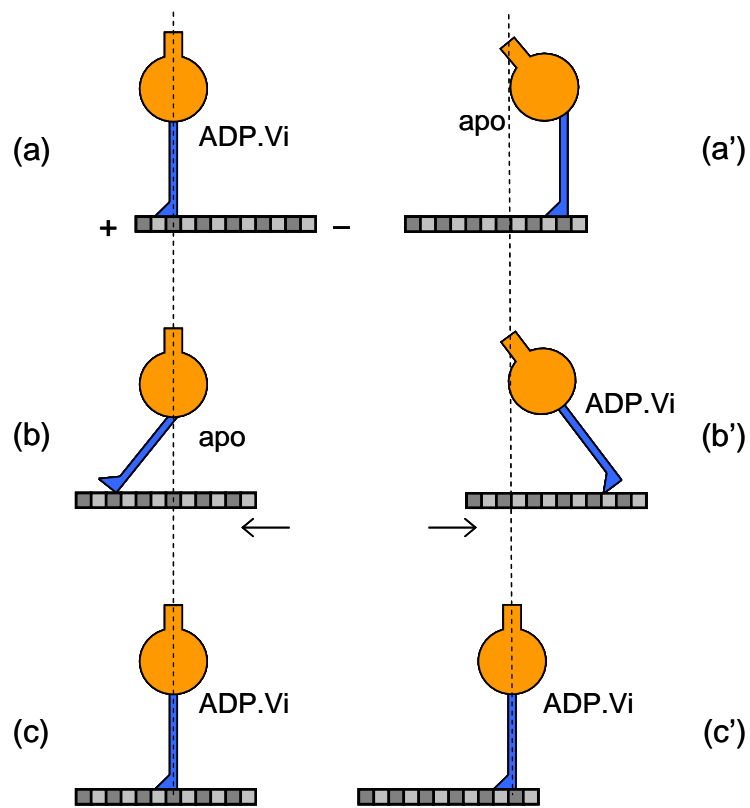

Figure 2. MT sliding by a fixed dynein. Strong MT-binding of the stalk tip occurs when the stalk is perpendicular to MT that is schematically shown in gray.

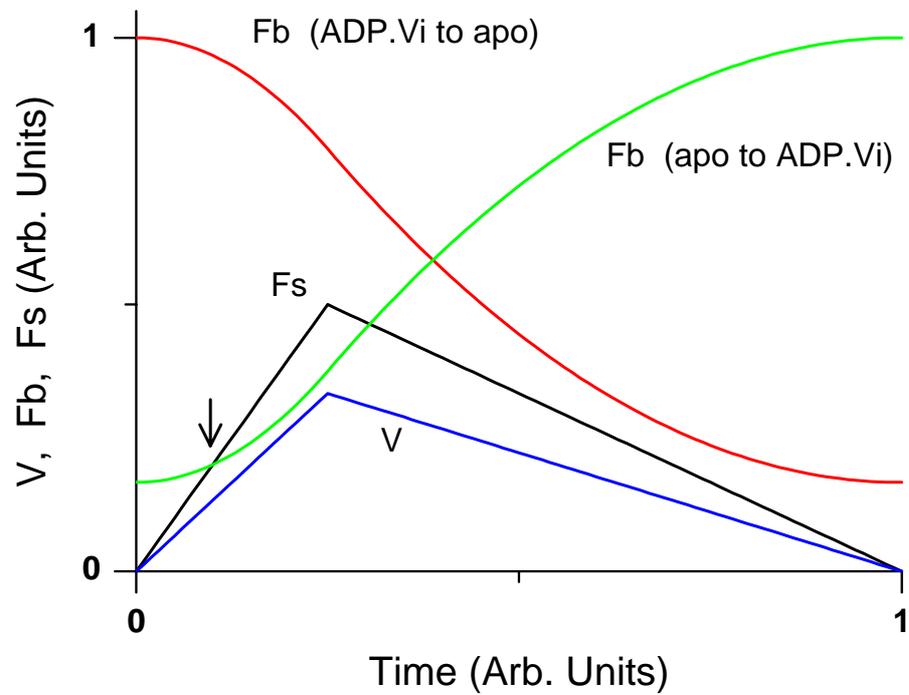

Figure 3. Temporal evolution of the rotation velocity V of the stalk, of the Stokes force $F_s$ on MT due to its movement driven by the stalk, of the MT-binding force $F_b$ of the stalk tip, during the rotation of the stalk as the nucleotide-state of dynein changes. The moment when $F_b$ becomes equal to $F_s$ is indicated by an arrow.

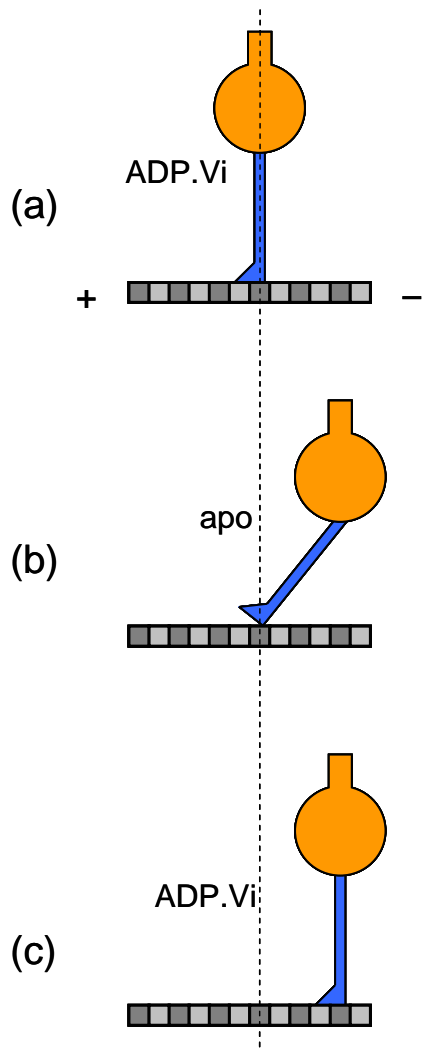

Figure 4.  Movement of dynein along a fixed MT.

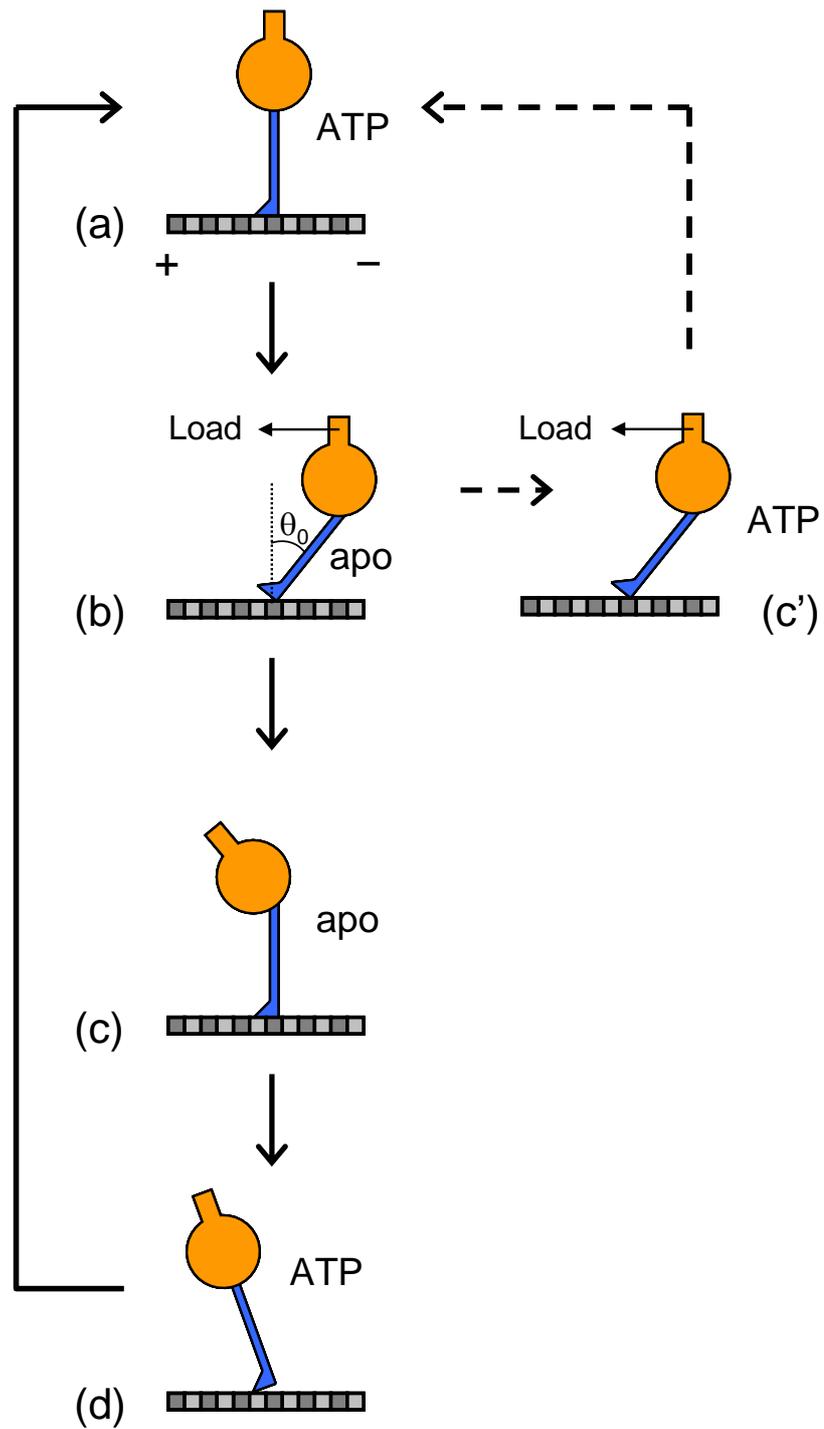

Figure 5. Movement of dynein along a fixed MT when there is a load acting on its head. The solid arrow lines show the pathway at low [ATP] and the dashed arrow lines show the pathway at saturating [ATP].